\documentstyle[aps,epsf,preprint]{revtex}

\def\1{{\bf 1}}
\def\[{\left[}
\def\]{\right]}
\def\be{\begin{eqnarray}}
\def\ee{\end{eqnarray}}
\def\nn{\nonumber}
\def\({\left(}
\def\){\right)}

\def\labels#1{\label{#1}}
\def\eq#1{(\ref{#1})}

\begin{document}
\def\h{{1\over 2}}
\def\s{\sigma}
\def\l{\lambda}
\def\v#1{\vec #1}
\def\.{\cdot}
\def\e{\epsilon}

\title{Consequences of a $Z_2$ Symmetry for Neutrino Oscillations}
\author{C.S. Lam}
\address{Department of Physics, McGill University\\
 3600 University St., Montreal, QC, Canada H3A 2T8\\
Email: Lam@physics.mcgill.ca}
\maketitle

\begin{abstract}
A $Z_{2L}\times Z_{2R}$ generation symmetry in the neutrino 
sector predicts 
the atmospheric neutrino mixing
to be maximal, and the MNS matrix element $U_{e3}$ to be zero,
consistent with observations.
Solar neutrino mixing may be maximal but is not required 
by the symmetry.
Neutrino masses of the first two generations
are predicted to vanish,  
providing a first approximation to
the oscillation data. The consequence of a smaller $Z_2$
symmetry is also discussed. In that case, deviation from the 
$Z_{2L}\times Z_{2R}$ result is of the order of the neutrino
mass ratio between the first two generations and the third generation.
\end{abstract}

\bigskip\bigskip
Right-handed neutrinos must be present to explain the deficit of solar and
atmospheric neutrinos through oscillations.
These are unique probes at high energy for they have
no standard-model quantum numbers to allow a
pollution by standard-model
interactions.
Recent results from Super-Kamiokande suggest that the mixing for
solar neutrinos and the atmospheric neutrinos are both maximal,
and the $(\Delta m)^2$ for solar neutrino oscillation is much smaller
than the $(\Delta m)^2$ for atmospheric neutrino oscillations \cite{SK}.
The negative result of CHOOZ \cite{CH} also limits the magnitude of the MNS
mixing matrix \cite{MNS} element
$|U_{e3}|^2$ to be smaller than 0.02 to 0.047, depending on the
exact value of the atomospheric neutrino mass
difference. Is there a simple way to understand
these facts?
Many models have been proposed \cite{MO}.
We suggest that a $Z_{2L}\times
Z_{2R}$ generation symmetry in the sea-saw scenario
will naturally explain many of these
observed results. By $Z_2$ we mean a finite group with elements
$\{e,g\}$ so that $g^2$ is the identity element $e$. 
Specifically, under such a symmetry, the atmospheric
neutrino mixing is maximal, and the MNS matrix element $U_{13}\equiv
U_{e3}$ is zero.
The neutrino masses of the first two generations vanish, which may be
a fair approximation to Nature since the mass difference observed
in atmospheric neutrino oscillation is much larger than that seen in
solar neutrino oscillation. Under such a symmetry, the solar neutrino
mixing may be maximal but is not required to be so.

Let $D$ be the Dirac mass matrix for neutrinos in the flavor
basis, and
$M$ the Majorana mass matrix in the same basis
for right-handed neutrinos. We shall assume oscillations to occur
only among the active neutrinos, so
both are $3 \times 3$ complex matrices, although $M$
is symmetrical. When the right-handed neutrinos are integrated out,
the effective mass matrix for the left-handed neutrinos 
in the flavor basis is \cite{AK}
\be
m'=D^TM^{-1}D.\labels{md}\ee
To diagonalize it, we must use the MNS matrix $U$ to rotate the flavor
basis into the energy eigenbasis, then
\be
m'=U^*mU^\dagger,\labels{mu}\ee
with $m={\rm diag}(m_1,m_2,m_3)$. 

Let $\nu_{aL}$ and $\nu_{aR}$ be respectively the 
left- and right-handed neutrino field
of the $a$th generation. Assume the mass terms in the neutrino
Lagrangian to be invariant under the
$Z_{2L}\times Z_{2R}$ transformation $\nu_{2A}\to -\nu_{3A},
\ \nu_{3A}\to -\nu_{2A}$, and $\nu_{1A}\to -\nu_{1A}$, where
$A$ is either $L$ or $R$.
Then the matrices $D$ and $M^{-1}$ must be of the form
\be
D&=&m_D\pmatrix{0&0&0\cr 0&1&-1\cr 0&-1&1\cr},\nn\\ \nn\\
M^{-1}&=&m_M^{-1}\pmatrix{\mu_1&\mu_2&\mu_2\cr \mu_2&1&\mu_3\cr \mu_2&\mu_3&1\cr}.\labels{dm}
\ee
Using \eq{md}, this leads to
\be
m'={2{m_D}^2(1-\mu_3)\over m_M}\pmatrix{0&0&0\cr 0&1&-1\cr 0&-1&1\cr}.
\labels{mp0}\ee
Other than the common scale factor which may be complex, the 
matrix is real and symmetrical, so it can be diagonalized by
a real orthogonal transformation. The eigenvalues are $(0,0,1)$,
but the matrix $U$ used for diagonalization is not unique because
of degeneracy of the first two eigenvalues. For the 
purpose of later generalization it is useful to deduce these
results explicitly from
\be
(m)_{ij}&=&U_{ai}(m')_{ab}U_{bj}\nn\\
&=&{2{m_D}^2(1-\mu_3)\over m_M}\(U_{2i}-U_{3i}\)\(U_{2j}-U_{3j}\).
\labels{m0}\ee
In order for $m$ to be diagonal, two of the following three 
combinations must vanish: $U_{21}-U_{31}, U_{22}-U_{32}$,
and $U_{23}-U_{33}$. Let us first consider the case when
$U_{23}-U_{33}\not=0$, but the other two combinations vanish.
In that case $m_1=m_2=0$ and $m_3=(2{m_D}^2(1-\mu_3)/m_M)
{(U_{23}-U_{33})}^2$, thus giving rise to the extreme limit
of a normal hierarchy for the neutrino masses, which is 
a good approximation to reality because $(\Delta m)^2_{12}\ll
(\Delta m)^2_{23}$. To simplify writing let $a=U_{21}
=U_{31}$,  $b=U_{22}=U_{32}$. Then $U$ is of the form
\be
U=\pmatrix{c&d&e\cr a&b&f\cr a&b&g\cr}.\labels{u}\ee
We will adopt the usual phase convention so that $c,d,f$ and $g$ are real,
and the imaginary part of $a$ and of $b$ are proportional to that of $e$.
Normalization of the second and third rows of
the matrix $U$ requires $|a|^2+|b|^2+f^2=1=|a|^2+|b|^2+g^2$.
Hence $f$ and $g$ have the same magnitude.
Orthogonality of the second and the third rows of $U$ then requires
$f=-g$ and $f^2=|a|^2+|b|^2=\h$.
In other words, atmospheric neutrino mixing is {\it maximal},
consistent with the Super-Kamiokande observation.
Using the fact that the first row of $U$ must be orthogonal to both
the second and the third roles, we conclude that $e=0$,
consistent with the CHOOZ observation.  In that case 
there will
be no CP violation observable through neutrino oscillations.
With the present phase convention
both $a$ and $b$ become real, and the magnitudes of $a,b,c,d$
are related by unitarity. So, there is only one free parameter $\theta_{12}$
left in describing $U$. For $c=-d=1/\sqrt{2}$, maximal mixing occurs
in solar neutrino mixing, but this is not required 
to be so by the symmetry. 
In summary, judging from
neutrino oscillations, the $Z_{2L}\times
Z_{2R}$ symmetry appears to be a good approximation to Nature. 

The other two solutions of \eq{m0} leads to either $m_1\not=0$
or $m_2\not=0$, with the other two diagonal matrix elements of
$m$ zero. The former case leads to $U_{11}=0$, and the latter
case leads to $U_{12}=0$, neither is consistent with solar neutrino
and CHOOZ observations. They will therefore be rejected.

Note also that the symmetry interchanges generations 2 and 3, with a 
minus sign. If instead it interchanged generations 1 and 2, which at first
sight might seem a more reasonable thing to do in view of the
mass hierarchy, the result would be the same as above but with the 
first and third rows of the matrix $U$ permuted. This would not be
consistent with experiment.

If the charged lepton and the quark Dirac mass matrices are subject
to the same symmetry, then all of them would have a mass spectrum
proportional to (0,0,1), not a bad first approximation. In that
case the CKM matrix would be $\1$, again a reasonable first approximation
given that the Cabibbo angle is small.

There is the question of how a small but non-zero
value of $m_{1,2}$ can be obtained. 
One possibility is to have a small breaking of
$Z_{2L}\times Z_{2R}$, into a diagonal $Z_2$ where the left-handed
and the right-handed particles are simultaneously transformed.
This does not alter the parametrization of $M$, but allows small
parameters to occur in $D$ where its present matrix elements
are zero. It should be possible to tune these parameters
to get finite masses for the first two generations. Since a different
set of parameters may be present for the charged leptons, the 
up-type quarks, and the down-type quarks, it is quite conceivable
that different mass spectra can be obtained for these different
particles, while obtaining also a realistic CKM matrix.

Let us now explore this scenario
more systematically. The $Z_2$ symmetry requires the (symmetric) matrix
$m'$ in \eq{md} and \eq{mu} to obey $m'_{12}=m'_{13}$ and
$m'_{22}=m'_{33}$. Using \eq{mu}, these two requirements can be
written as
\be
\sum_i\alpha_i\beta_im_i&=&0,\labels{m1}\\
\sum_i\gamma_i\beta_im_i&=&0,\labels{m2}\ee
where
\be
\alpha_i&=&U^*_{1i},\nn\\
\beta_i&=&U^*_{2i}-U^*_{3i},\nn\\
\gamma_i&=&U^*_{2i}+U^*_{3i},\nn\ee
thus giving rise to constraints on $U$ controlled by the neutrino
mass values. For the extreme hierarchy spectrum
$(m_1,m_2,m_3)\propto(0,0,1)\equiv s_1$ discussed above, the constraints are
$\alpha_3\beta_3=\gamma_3\beta_3=0$. Being second order equations
there are two solutions: (1a) 
$\alpha_3=\gamma_3=0$, and (1b) $\beta_3=0$. Solution (1a) is the same
as the $Z_{2L}\times Z_{2R}$ solution before. To see that, note that
the third column of $U$ is fixed by this constraint 
and unitarity to be 
$(U_{13},U_{23},U_{33})=(0,f,-f)$, with $f^2=\h$. Orthogonality
of the first two columns to the third  then fixes $U$
to be of the form \eq{u}, with parameters identical to those in
$Z_{2L}\times Z_{2R}$ as per the arguments given below \eq{u}.
Solution (1b) gives $U_{23}=-U_{33}$, which is not very restrictive.
For a realistic hierarchical
spectrum $s_2=(\eta,\epsilon,1)$, with $\epsilon\ll 1$ and $\eta$
either of the same order or much less than $\epsilon$, the constraints
become $\alpha_3\beta_3=O(\epsilon)$ and $\gamma_3\beta_3=O(\epsilon)$.
The two corresponding solutions are: (2a) $\alpha_3=O(\epsilon)$,
$\gamma_3=O(\epsilon)$, and $\beta_3=O(1)$, and (2b) $\beta_3=O(\e)$
and $\alpha_3,\gamma_3=O(1)$. As before, solution (2b) is not very 
restrictive.
For (2a), $|U_{e3}|$
is of order $\e$ and the atmospheric neutrino oscillation is close to maximal,
with deviation of the order of $\e=m_2/m_3$. In this way a $Z_2$ symmetry
with a realistic mass spectrum as an input gives rise to practically the
same predictions as the larger $Z_{2L}\times Z_{2R}$ symmetry.

Whether a more realistic mass spectrum  should be obtained by an explicit
breaking of $Z_{2L}\times Z_{2R}$ like the one described above,
by renormalization-group corrections, or by something else, will
be left to future investigations.

This research is supported in part by the Natural Sciences and
Engineering Research Council of Canada, and the Fonds pour la
formation de Chercheurs et l'Aide \`a la Recherche of Qu\'ebec.
I am indebted to T.K. Kuo, Greg Mahlon, Gary Shiu, and Tony Zee
 for stimulating discussions.


\begin{thebibliography}{9}
\bibitem{SK} Y. Suzuki, `Solar neutrino results from Super-Kamiokande',
H. Sobel, `The study of atmospheric neutrino with Super-Kamiokande',
Neutrino 2000 Conference (Sudbury, Canada; June, 2000)
\bibitem{CH} CHOOZE Collaboration, M. Apollonio {\it et al}., hep-ex/9907037.
\bibitem{MNS} Z. Maki, M. Nakagawa, and S. Sakata, {\it Prog. Theor. Phys.}
{\bf 28} (1962) 870.
\bibitem{MO} V. Barger, S. Pakvasa, T.J. Weiler, and K. Whisnant, 
hep-ph/9806387; A.J. Baltz, A.S. Goldhaber, and M. Goldhaber,
hep-ph/9806540; 
 Y. Nomura and T. Yanagida, hep-ph/9807325;
G. Altarelli and F. Ferugio, hep-ph/9807353, 9809596, 9812475;
S. Raby, hep-ph/9909279; 
R.N. Mohapatra and S. Nussinov, hep-ph/9807415;
S.K. Kang and C.S. Kim, hep-ph/9811379;
R. Barbieri, L.J. Hall, G.L. Kane, and G.G. Ross, hep-ph/9901228;
E. Ma, hep-ph/9902392, 9909249;
R.N. Mohapatra. A. P\'erez-Corenzana, and C.A. de S. Pires,
hep-ph/9911395;
T.K. Kuo, G.-H. Wu, and S.W. Manosour, hep-ph/9912366;
D. Chang and A. Zee, hep-ph/9912380;
K.R.S. Balaji, A.S. Dighe, R.N. Mohapatra, and M.K. Parida,
hep-ph/0001310; A. Aranda, C.D. Carone, and R.F. Lebed, hep-ph/0002044;
C.H. Albright and S.M. Barr, hep-ph/0003251;
Y. Koide and A. Ghosal, hep-ph/0008129.
\bibitem{AK} See, for example, 
G. Altarelli and F. Feruglio, hep-ph/9905536, and
E.Kh. Akhmedov, hep-ph/0001264, for excellent reviews.  
\end{thebibliography}
\end{document}